\documentstyle[12pt,epsfig]{article}
\def\ltsim{\lower3pt\hbox{$\, \buildrel < \over \sim \, $}}
\def\gtsim{\lower3pt\hbox{$\, \buildrel > \over \sim \, $}}
\def\be{\begin{equation}}
\def\ee{\end{equation}}
\def\ba{\begin{eqnarray}}
\def\ea{\end{eqnarray}}
\def\ga{\mathrel{\raise.3ex\hbox{$>$\kern-.75em\lower1ex\hbox{$\sim$}}}}
\def\la{\mathrel{\raise.3ex\hbox{$<$\kern-.75em\lower1ex\hbox{$\sim$}}}}

\newcommand{\bi}[1]{\bibitem{#1}}

\begin{document}
\baselineskip=16pt
\begin{titlepage}
\rightline{OUTP-00-08P}
\rightline{hep-th/0003074}
\rightline{March 2000}  
\begin{center}

\vspace{0.5cm}

\large {\bf Brane Universe and Multigravity: 
Modification of gravity at large and small distances.}

\vspace*{5mm}
\normalsize

{\bf Ian I. Kogan\footnote{i.kogan@physics.ox.ac.uk}
 and Graham G. Ross \footnote{g.ross@physics.ox.ac.uk}}

\smallskip 
\medskip 
{\it $^a$Theoretical Physics, Department of Physics, Oxford University}\\
{\it 1 Keble Road, Oxford, OX1 3NP,  UK}
\smallskip

\vskip0.6in \end{center}
 
\centerline{\large\bf Abstract}

We consider a modification of gravity at large distances in a Brane
Universe which was discussed  recently \cite{Oxford} - \cite{DGP}. 
In these models the modification of gravity at large distances is ultimately
connected to existence of negative tension brane(s) and  exponentially 
small tunneling factor. We  discuss a general model which interpolates
 between Bi-gravity $+-+$ model \cite{Oxford} and GRS model
\cite{GRS}. We  also  discuss   the possible mechanism of
stabilization  of negative tension branes in AdS background.
Finally we show that extra degrees of freedom  of massive  gravitons
 do not lead to disastrous contradiction with General
Relativity if the stabilization condition 
$\int dy\,\sqrt{-G^{(5)}}\,(T^\mu_\mu-2T^5_5)=0$ 
\cite{kkop1} is implemented.

\vspace*{2mm} 

\end{titlepage}

Recently, there has been considerable interest in theories in which the
SM fields are localized on a 3-brane in a higher dimensional spacetime. 
 The  idea that  a multidimensional Universe may not be of KK 
type, but rather a low dimensional ($3+1$ in the case of our Universe)
brane  dates  back to early eighties when  independently
 Akama \cite{akama} and Rubakov and Shapochnikov \cite{rubshap} suggested
models of our Universe as topological defects and later Visser
\cite{visser} and Squires \cite{squires} described how particles can
be  gravitationally trapped on the brane. The  possibility that gravity
 be trapped on a brane was  suggested by Gogberashvili \cite{gog}

In  papers of Antoniadis, Arkani-Hamed, Dimopoulos and Dvali 
\cite{D1,D2,ad1} it was  shown that the size of extra dimensions 
 can be large, up to a mm scale. Then   Randall and Sundrum proposed
\cite{RS} a  scenario in the  geometry is
non-factorizable  and even infinitely large fifth dimension leads to
the Newton law at macroscopic distances along the brane. In this
picture the concept of a massless graviton localized on a positive
tension brane was of utmost importance. Another
important element was a negative tension brane. 

Then in was suggested in  \cite{Oxford} and later and independently in 
\cite{GRS}  that in some modification of RS scenario one can have even 
more strange behaviour - namely gravity is modified at both large and
small scales. Even if explicit realizations of this idea suggested  
by Oxford group (KMPRS) \cite{Oxford} and Gregory, Rubakov and 
Sibiryakov (GRS) \cite{GRS}  are   different  one can  
see that they actually related to each other and share several key
elements - for example the presence of  free negative tension branes.
There is another common  feature in both models - exponential
suppression and tunneling effect (in case of GRS model it was
discussed by Csaki, Erlich and  Hollowood in \cite{CEH}).
In this letter we demonstrate that these two models are actually 
two limiting cases of a more general  model
which we  shall call $+--+$ or ``multigravity'' model and which
interpolates between ``bi-gravity'' \cite{Oxford} and ``quasi-localized''
 gravity \cite{GRS}, \cite{CEH}.
 Besides this we are going to discuss the mechanism which 
will prevent negative brane destabilization - which is crucial for all 
these models. At the end we are going to discuss an important
observation of Dvali, Gabadadze and Porrati \cite{DGP} that in this
models four-dimensional graviton is massive and this leads to discrepancy 
 \cite{VZ} between Newton law and light bending  (and  other
experimentally observed post-Newtonian effects, for example  Mercury
perihelion precession). The remarkable resolution of this potentially 
disastrous discrepancy is that in a full theory there is actually a
smooth limit for graviton mass $m \rightarrow 0$  if one take into
account  constraint $\int dy\,\sqrt{-G^{(5)}}\,(T^\mu_\mu-2T^5_5)=0$
  which is a  necessary condition for 
 stable compactification and correct  cosmology on a brane \cite{kkop1}

\begin{figure}[ht]
\begin{center}
\mbox{\epsfig{file=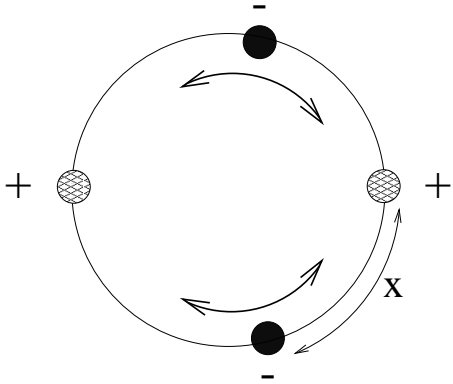,width=6cm}}
\caption { $+-+$ model with two $''+''$ branes at the fixed points and moving
$''-''$ branes. In the limiting case when $x \rightarrow 0$ we have a RS 
configuration.
\label{Fig.1}}
\end{center}
\end{figure}

Let us first discuss the KMPRS model \cite{Oxford} which consists 
of two positive  branes located at the fixed points of a 
$S_1/Z_2$ orbifold with one negative brane which can move freely in
 between (see 
Fig.~\ref{Fig.1}).
 It is easy to see that the two-brane RS model is nothing but the limiting
 case of our three-brane model, when the negative brane hits one of the 
positive branes.The model has two parameters, 
(at a fixed  bulk curvature) - the warp factor and the $x$ 
factor - which  effectively measures the distance between one of the positive
branes and the negative brane. The  RS model corresponds to the
limiting  case $x = 0$.

The GRS model can be obtained from this model by cutting negative
brane in half, i.e. instead of one  $''-''$ brane with tension $-\Lambda$
 one can take two branes with negative tension $-\Lambda/2$ each and
then move them apart, as on Fig.2 If one then move the second $''+''$
 brane to infinity together with one of the   $''-1/2''$ branes we
shall get precisely the GRS picture - and the space between  $''-1/2''$
brane is, of course, flat. Thus we have  a new model - the 
$+--+$ model which depends on 3 parameters - $l_L$ and $l_R$ which are 
the  differences between coordinates of left $''+''$ and $''-1/2''$
branes and right
$''+''$ and  $''-1/2''$ branes, and $l_{--}$ which is a difference
between  coordinates of the two  $''-1/2''$ branes.
This model  consists of four parallel 3-branes in an $AdS_5$ space with
cosmological constant $\Lambda<0$. The 5-th dimension has the geometry
of an orbifold and the branes are located at
$y_0=0$, $y_1 = l_L$ and $y_2 = l_L+ l_{--}$ and $y_3 =y_2 + l_R$,
 where $y_0$ and $y_2$
are the orbifold  fixed
points\footnote{The requirement that we have orbifold fixed points is
not really necessary for our analysis, which is much more general}
 (see Fig.2). 
Firstly we consider the branes having no  matter on them in
order to find a suitable vacuum solution. The action of this setup is:
\be
S=\int d^4 x \int_{-y_2}^{y_2} dy \sqrt{-G} 
\{-\Lambda + 2 M^3 R\}-\sum_{i}\int_{y=y_i}d^4xV_i\sqrt{-\hat{G}^{(i)}}
\ee
where $\hat{G}^{(i)}_{\mu\nu}$ is the induced metric on the branes
and $V_i$ their tensions. Here we  included  negative $y$ and we are
looking for solutions invariant with respect to $Z_2$ symmetry $y
\rightarrow -y$. The notation is the same as in
Ref. \cite{RS}.

At this point we demand that our metric respects 4D Poincare\'{e}
invariance. The metric ansatz with this property is the following:
\be
ds^2=e^{-2\sigma(y)}\eta_{\mu\nu}dx^\mu dx^\nu +dy^2
\ee
Here the ``warp'' function $\sigma(y)$ is essentially a conformal
factor that rescales the 4D component of the metric. 
It satisfies the following differential equations:
\ba
\left(\sigma '\right)^2&=&k^2, ~~~~~~~
\sigma ''= \sum_{i}\frac{V_i}{12M^3}\delta(y-L_i)
\ea
where $
k=\sqrt{\frac{-\Lambda}{24M^3}}$ is a measure of the curvature of the bulk. 
The brane tensions are tuned to $V_0=-\Lambda/k>0$,
$V_1= V_2 = \Lambda/2k<0$, \mbox{$V_2=-\Lambda/k>0$}. 
It is convenient to introduce 3  dimensionless parameters
\ba
x_L = kl_L, ~~~ x_R = k l_R, ~~~ x_{-} =   k l_{--}
\ea
\begin{figure}[ht]
\begin{center}
\mbox{\epsfig{file=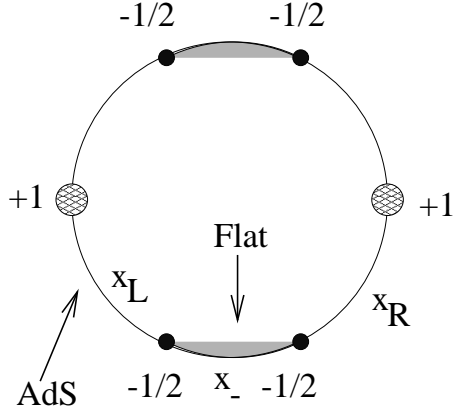,width=6cm}}
\caption{GRS model can be obtained from the left part of
  $+--+$ configuration. The whole configuration can be
 considered as  two GRS models connected via flat space.
\label{Fig.2}}
\end{center}
\end{figure}
It is easy to see that when $x_{-} =0$ we have $+-+$ model \cite{Oxford}
with the hierarchy factor $w = e^{(x_R-x_L)}$ and  parameter $x$
 on Fig.1 is $x_R$. In the limit of infinite $x_{-}$ we have two
infinitely separated GRS systems.  Then it becomes quite clear that
 $+--+$ model smoothly interpolates between this two models.
spectrum that follows from the dimensional reduction.
 Then we have to 
 find the spectrum of  (linear) fluctuations of the metric:
\be
ds^2=\left[e^{-2\sigma(y)}\eta_{\mu\nu} +\frac{2}{M^{3/2}}h_{\mu\nu}
(x,y)\right]dx^\mu dx^\nu +dy^2
\ee

We expand the field $h_{\mu\nu}(x,y)$ in graviton and KK states plane waves:
\be
h_{\mu\nu}(x,y)=\sum_{n=0}^{\infty}h_{\mu\nu}^{(n)}(x)\Psi^{(n)}(y)
\ee
where
$\left(\partial_\kappa\partial^\kappa-m_n^2\right)h_{\mu\nu}^{(n)}=0$
and fix the gauge as
$\partial^{\alpha}h_{\alpha\beta}^{(n)}=h_{\phantom{-}\alpha}^{(n)\alpha}=0$.
The function $\Psi^{(n)}(y)$ will obey a second order differential
equation which after a change of variables reduces to an ordinary
Schr\"{o}dinger equation:
\be
\left\{-
\frac{1}{2}\partial_y^2+V(y)\right\}\hat{\Psi}^{(n)}(y)=
\frac{m_n^2}{2}\hat{\Psi}^{(n)}(y), ~~~~
\hat{\Psi}^{(n)}(y)\equiv \Psi^{(n)}(y)e^{\sigma/2}
\ee
where potential $V(y)$  is determined by $\sigma(y)$. Qualitatively it is
 $\delta$-function potentials (attractive for $''+''$ and repulsive
 for $''-''$ branes) of
different weight depending on brane tension  and an extra smoothing
term  (due to the AdS geometry) 
that gives the
potential a  ``volcano'' form. 

An interesting characteristic of this potential is that it always
gives rise to a (massless) zero mode which reflects the fact that
Lorentz invariance is preserved in 4D spacetime.  This mode is
normalizable for finite $x_{-}$ and becomes non-normalizable in a GSR limit
 of infinite $x_{-}$.
The interaction of the linearized gravitons  to  matter localized on a 
brane located at some  $y$ is given by 
\ba
{\mathcal{L}}_{int}=\frac{f(y)}{M^{3/2}}\sum_{n\geq 0}
\Psi^{(n)}(y)h_{\mu\nu}^{(n)}(x)T_{\mu\nu}(x) 
\ea
with $T_{\mu\nu}$ the energy momentum tensor of the SM Lagrangian and
 $f(y)$ is some universal function. From this expression the Newton
potential  on a brane is given by
\be
U(r) \sim  \sum_{n\geq 0} \frac{(\Psi^{(n)})^2(y)}{M^3} \frac{e^{-mr}}{r}
~~ \sim \int dm  \frac{\Psi_{m}^2(y)}{M^3} \frac{e^{-mr}}{r}
\ee
where $\rho(m)$ is the spectral density and it is discrete for $x_{-}=0$
and becomes continuous in a  GRS limit of infinite $x_{-}$.
We shall present full solution of this model for all $x_{-}$ elsewhere 
\cite{gang4}.

Let us remind that one of the striking predictions of  $+-+$ model
 is the fact that the first KK mode can be very light and strongly 
coupled compared to the rest of the KK states. 
 This light mode can
be so light  that the corresponding wavelength
 can be by order of $1\%$ of the observable size of the Universe while
 the {\it second} KK mode is in submillimeter region. 
Surprisingly enough this situation is not excluded experimentally and 
we called it  ``Bi-Gravity'' -  in all
experimentally analyzed regions gravitational attraction is due to an
exchange of two particles - the massless graviton and ultralight  first KK 
mode. Only at scales larger than $10^{26}cm$ will the first KK mode decouple 
leading to a  smaller gravitational coupling beyond this length scale. 
The reason the anomalously light KK mode exists is due to the fact that with 
more that one
$''+''$ brane there will be a bound state on {\it each} of them when they have
infinite separation. At finite distances there is a mixing between the two 
localized states. One superposition  is the true ground
state while the other
configuration has non-zero mass, but the gap may be very small - it
 is given by a tunneling factor. 
 In the case $x>>1$ we obtained a reliable
approximation to its mass 
\be
m_1=2ke^{-x_L-x_R} 
\label{m1}
\ee
For $w =1$  we get $m_1=2ke^{-2x}$ and 
 the wave functions  $ \Psi_{0}^2(0) = \Psi_{1}^2(0)$ with 
exponential accuracy. The 
masses of the other KK states in the above region are found to
depend in a different way on the parameter $x$. The mass of the
second state and the spacing $\Delta m$ between the subsequent states have the 
form:
\ba
m_2 \approx k e^{-x}, ~~~
\Delta m \approx \varepsilon k e^{-x}
\label{m2}
\ea
where $\varepsilon$ is a number between 1 and 2. 
Equations (\ref{m1}) and (\ref{m2}) show that, for large $x$,
 the lightest KK mode 
splits off from the remaining tower. This leads to an exotic possibility in 
which the lightest KK mode is the dominant source of Newtonian gravity!

Cavendish experiments and astronomical observations studying the motions of 
distant galaxies have put Newtonian
gravity to test from submillimeter distances up to distances that
correspond to $1\%$ of the size of observable 
Universe, searching for violations of the weak equivalence principle
and inverse square law. In the context of the graviton KK modes discussed above 
this constrains $m<10^{-31}{\rm eV}$ or $m>10^{-4}{\rm eV}$. Our exotic scheme corresponds 
to the choice $m_1\approx 10^{-31}{\rm eV}$ and $m_2>10^{-4}{\rm eV}$. In this case, for 
length scales less than $10^{26}{\rm cm}$ gravity is generated by the exchange of {\it 
both} the massless graviton and the first KK mode (C is some constant)
\be
U(r) = C \frac{(\Psi^{(0)})^2(0)}{M^3} \left(\frac{1}{r} +
  \frac{(\Psi^{(1)})^2(0)}{(\Psi^{(0)})^2(0)} \frac{e^{-m_1 r}}{r}\right) +
 O(e^{-m_2 r})
 \approx  2C \frac{(\Psi^{0})^2(0)}{M^3} \frac{1}{r}
\ee
so gravitational constant is $G_N =  2C \frac{(\Psi^{0})^2(0)}{M^3}$.
According to this picture deviations from Newton's law will appear
in the submillimeter regime $ m_2 r < 1$ as the 
Yukawa corrections of the second and higher KK
states become important.  Also the
presence of the ultralight first KK state will give deviations
from Newton's law as we probe cosmological scales $m_1 r >1$ 
(of the order of the observable universe) with $G_N/2$ instead of
$G_N$. The phenomenological signature 
of this scenario is that gravitational
interactions will appear to become weaker (in general case by the
factor  $w$ \cite{Oxford}) for distances larger than $1/m_1 \sim
 10^{26}{\rm cm}$!

\begin{figure}[ht]
\begin{center}
\mbox{\epsfig{file=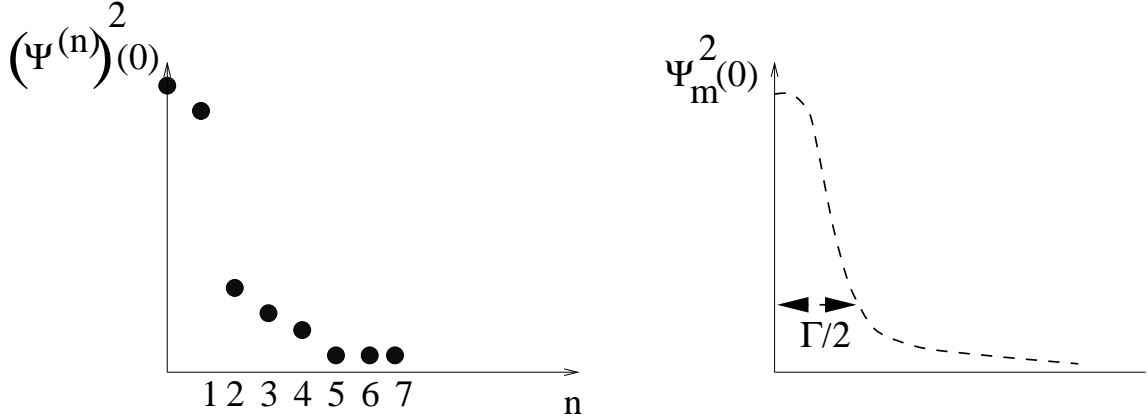,width=16cm}}
\caption{Behaviour $\Psi_{m}^2(0)$ in a $+-+$ model (discrete
spectrum). For a general $+--+$
model the discrete spectrum  density increases with the increase of
$x_{-}$ and spectrum becomes continuous in  GRS limit.
\label{Fig.3}}
\end{center}
\end{figure}

The idea behind GRS construction was totally different - they did not
have normalized modes, but rather continuous spectrum, however they
have a ``resonance'' \cite{CEH} which effectively was due to the fact that 
that the negative brane created a tunneling factor (negative brane
acts as a repulsive potential)  which  effectively leads to  a
resonance in a wave function $\Psi_{m}(0)$ 
of gravitons
\be
\Psi_{m}^2(0) = \frac{c}{m^2 + \Gamma^2/4} + O(m^4)
\label{resonance}
\ee

From  (\ref{resonance}) one can  see \cite{GRS}, \cite{CEH}
\be
U(r) = \frac{2c}{M^3 \Gamma} \frac{1}{r} \int \frac{dx}{x^2+1}
\exp\left(-\frac{\Gamma r}{2} x\right)
\ee
and we see that for $r << 1/\Gamma$ we recover the Newton $1/r$
potential with 
\be
G_N =   \frac{\pi c}{M^3 \Gamma} 
\ee
 and for $r > 1/\Gamma$   potential is $1/r^2$ and gravity  is modified 
at large distances.

But this is just the same as we have on our discrete case (see Fig.3)
 - the reason is of course that $\Gamma \sim e^{-x}$.  Thus we see that in
both limits it the same tunneling factor $e^{-x}$ made modification of 
gravity possible - in discrete case it is exponentially small
splitting between masses of massles graviton and first KK mode, in
continuous case it is exponentially small virtual resonance width.
\begin{figure}[ht]
\begin{center}
\mbox{\epsfig{file=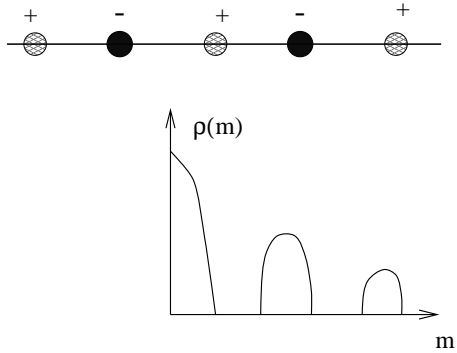,width=6cm}}
\caption{A periodic array of positive and negative branes is a model
of Crystal Brane Universe \cite{kaloper}. 
\label{Fig.4}}
\end{center}
\end{figure}

Of course one get the same behaviour in other models when  positive
branes are separated by negative branes. For example  one may have a
crystal \cite{kaloper} and in this case we have bands. Potential will
be proportional to
\be
U(r) 
~~ \sim \int dm \rho(m)  \frac{e^{-mr}}{r}
\ee
where $\rho(m)$ is band density and for narrow band we shall have
multigravity again. All we need is that the width of the first band
$\Gamma_1$ is much smaller than the separation between first and
second bands (more details will be given in \cite{gang4}).

 Let's make  some comments about the stability of the brane configuration
discussed here. It involves a negative tension brane located between
positive tension branes. Necessarily the negative tension (three) brane is
free to move (in this configuration it cannot be located at a fixed point)
and the question naturally arises about its stability. Being negative
tension it will be advantageous to increase its volume and this may happen
both by changing its location through the effects of the warp factor or
through the brane bending. It is instructive to consider the first effect on
its own by determining the effect of moving the position of the brane a
distance $\delta $ from an initial position midway between the positive
branes. In this case \cite{Oxford} the warp factor at the negative
brane decreases by $\ $the factor $e^{-\kappa \delta }$, where $\kappa
$ is the five dimensional curvature. Thus the volume of the negative brane
decreases, implying the midpoint is a position of equilibrium for the
``flat'' brane. Consider now the effect of deforming the brane over some
transverse size $l$ which may reasonable be taken to be the horizon size.
The mean displacement of the brane about its equilibrium position is
proportional to $R-R\cos \theta \sim R\theta ^{2}=l\theta ,$ where $\theta
=l/R\ll 1.$ Thus the warp factor decreases by the factor $e^{-2c\kappa
l\theta }$ where $c$ is some constant. The deformed brane stretches by an
amount $\approx $($1+\theta ^{2}/2)$ in each transverse dimension so the net
change in the brane volume is ($e^{-c\kappa l\theta }(1+\theta
^{2}/2))^{3}.$ Since the horizon size, $l,$ will always be much larger than
the Planck length $\approx \kappa ^{-1}),$ this naive argument suggests
that the warp factor is the dominant effect and that the bending of the
brane is energetically disfavoured. If this is true the brane configurations
discussed in this paper may be stable structures and the modified
gravitational interactions discussed above may indeed be possible .

Finally let us  discuss  problems raised in \cite{DGP}. Contrary to massless
 four-dimensional  graviton which has 2  polarizations the massive one 
has 5 possible polarizations. The tensor structure for graviton
propagator is given by
\be
G^{\mu\nu,\alpha\beta}= C\frac{\left[1/2\left(g^{\mu\alpha}g^{\nu\beta} +
g^{\nu\alpha}g^{\mu\beta}\right) - t g^{\mu\nu}g^{\alpha\beta}\right] 
 + O(pp/p^2) }{p^2 -m^2}
\ee
where for all non-zero m parameter $t = 1/3$ \cite{VZ} but for $m=0$
 $t = 1/2$. Then the  one-graviton exchange amplitude for any two 
four-dimensional sources $T^1_{\mu\nu}$ and $T^2_{\alpha\beta}$ is
given by
\be
T^1_{\mu\nu}G^{\mu\nu,\alpha\beta}T^2_{\alpha\beta} 
= C \frac{T^1_{\mu\nu}T^{2~\mu\nu} - t T^1T^2}{p^2-m^2}
\ee
where $T = T_{\mu}^{\mu}$.
Massive bodies  at rest  have only non-zero  $T_{00}$ and $T =
T_{00}$. Newton constant can be found  from amplitude
\be
\frac{C(1-t)}{p^2-m^2}T^1_{00}T^2_{00}
\ee
But there is another type of matter - light and for light $T = 0$.
The amplitude for light bending does not depend on $t$ and  is given
by
\be
\frac{C}{p^2-m^2}T^1_{00}T^2_{00}
\ee
The ratio of two amplitudes is $1-t$.
Because both amplitudes are known experimentally  it is
easy to find $t$ -and of course it must be $t=1/2$ as in Einstein
General relativity. If we take $t=1/3$ the ratio of two amplitudes is
 going to be $2/3$ instead of $1/2$ - and  this is factor $4/3$ which
 was discussed in \cite{DGP}.

 However this is not the whole story. We have a five-dimensional
 General relativity and the  graviton propagator in five dimensions
has tensor structure \cite{KK}
\be
G^{MN, PQ} \sim \left[1/2 \left(g^{MP}g^{NQ} + g^{MQ}g^{NP}\right) - 
 t g^{MN}g^{PQ}\right]  + O(pp/p^2) 
\ee
 where $t = 1/3$ i.e. as {\it massive} four-dimensional graviton.
 The reason is very simple - they both have 5 physical components.
 It is a well known fact  in KK theory  that after compactification 
the massless sector contains   massless graviton, graviphotn and KK
scalar -  in total five degrees of freedom. At the same time massive
KK excitations contain only massive 4d graviton.

 Let's then consider the full five-dimensional amplitude
\be
T_{MN}G^{MN, PQ}T_{PQ}  \sim
T_{MN} \left[1/2 \left(g^{MP}g^{NQ} + g^{MQ}g^{NP}\right) - 
 t g^{MN}g^{PQ}\right]T_{PQ}
\ee
where $T_{MN}$ is a full stress-energy tensor.
We don't have $T_{\mu 5}$ but besides $T_{\mu\nu}$ there is $T_{55}$.
Then the total amplitude can be written as
\ba
T_{\mu\nu}T^{\mu\nu} - t T T + (1-t) T_5^5  T_5^5 - 2t T  T_5^5 
 = \left[ T_{\mu\nu}T^{\mu\nu} - \frac{1}{2}T T \right] +
\nonumber \\
 (1/2-t) T T  + (1-t)   T_5^5  T_5^5  - 2t T  T_5^5
\ea
The first line is the correct four-dimensional amplitude - all known
experimental data gives this amplitude, so it will be nice to see that 
the second line is identically zero. Amusingly at $t = 1/3$ the second 
line is a full square
\be
\frac{1}{6} (T - 2 T_5^5)^2
\ee
Why it is natural to have the condition $(T - 2 T_5^5) = 0$ ?
 First of all this is written in momentum representation. All
amplitudes under consideration  have zero  momentum transfer in the
fifth direction. This means  that we have to  integrate over
 the fifth coordinate $\int dy\,\sqrt{-G^{(5)}}\,(T -2 T^5_5)$. But
 this is  precisely the integral condition which was discussed in 
\cite{kkop1}. The $g_{55}$ remains in equilibrium and 
the size of an extra dimension stays fixed if  the stress 
energy of the 5D matter satisfies the following constraint\footnote{This
constraint was also derived from a topological argument in \cite{E}.}:
\be
\int d^4x\,dy\,\sqrt{-G^{(5)}}\,(\hat{T}^\mu_\mu-2\hat{T}^5_5)=0.
\label{stable1}
\ee

To show  what does this condition  mean let us consider
 standard Kaluza-Klein decomposition
\be
G^{(5)}_{MN}= \left(\begin{tabular}{cc} $G^{(4)}_{\mu\nu}$ & $0$\\[1mm]
$0$ & $e^{2\gamma}$\end{tabular} \right)\,,
\label{kk}
\ee
and then consider the linear term 
\be
\delta G^{(5)}_{MN}\, T^{MN}=\delta G^{(4)}_{\mu\nu}\, T^{\mu\nu} +
\delta (e^{2\gamma})\,T^{55}
\ee
in {\em any} system. Using $G^{(4)}_{\mu\nu} = e^{-\gamma}
g_{\mu\nu}$, one can rewrite the above expression as
\be
\delta G^{(5)}_{MN}\, T^{MN} = 
e^{-\gamma} \delta g_{\mu\nu}\,T^{\mu\nu}
- \delta\gamma\,( T-2\hat T^5_5)
\ee
Thus, we see that $ T_{\mu\nu}$ is a source for the graviton while the
combination $(T-2T^5_5)$ is the source for the  scalar (radion).
In the presence of a mechanism that stabilizes the extra dimension
 (see for example \cite{GW} and \cite{randall3} and references
 therein) the source for the constant mode of this scalar is identically zero. 
If we do not have a stabilization mechanism but,
 nevertheless, we look for solutions with a Newtonian limit, 
we still have to require the absence of the above
term. 

 The conclusion is that  there is a smooth limit  $m \rightarrow 0$
 in multigravity. The situation here similar to the Higgs mechanism.
 There is no smooth limit when the mass of a vector particle is given
by hand - but there is a smooth limit when it emerges dynamically due
to spontaneous symmetry breaking.

{\bf Acknowledgments:} 
 We  would like to thank P. Kanti, S. Mouslopoulos, K.Olive,
A. Papazoglou, M. Pospelov and M. Voloshin for  interesting  discussions.
 This  work   is supported in part by PPARC rolling grant
PPA/G/O/1998/00567, the EC TMR grant FMRX-CT-96-0090 and  
 by the INTAS grant RFBR - 950567. One of us (I.I.K)  would like to
thank TPI, University of Minnesota for kind hospitality  and
stimulating discussions during visit in February 2000.

{\bf Note added in proof: }
 During preparation of  this paper for
publications  several papers were published in which the $4/3$ problem 
 \cite{DGP} was discussed \cite{last}, but not the constraint 
(\ref{stable1}).

\end{document}